\documentclass[
  aps,
  pra,
  reprint,
  superscriptaddress,
  amsmath,
  amssymb,
  longbibliography
]{revtex4-2}

\usepackage{graphicx}  
\usepackage{bm}        
\usepackage{amssymb}   
\usepackage{xspace}
\usepackage{verbatim}
\usepackage{color}
\usepackage{soul}
\usepackage{subfigure}
\usepackage[export]{adjustbox}
\usepackage{dcolumn}
\usepackage{amsthm,stmaryrd,mathtools}
\usepackage{txfonts}
\usepackage{braket}
\usepackage{amsmath}
\usepackage{xcolor}
\usepackage{array}
\usepackage{multirow}
\usepackage{tabularx}
\usepackage{booktabs}
\usepackage{lipsum}
\usepackage[normalem]{ulem}
\usepackage[colorlinks=true,linkcolor=blue,citecolor=blue,urlcolor=blue]{hyperref}
\usepackage[figure,figure*]{hypcap}
\usepackage{balance}

\newcommand{\appsection}[1]{%
  \section{\MakeUppercase{#1}}%
}

\begin{document}

\title{Quantum Metrology via Adiabatic Control of Topological Edge States}

\author{Xingjian He}%
\altaffiliation[]{These authors contributed equally to this work.}
\affiliation{Department of Physics, National University of Singapore, Singapore 117542, Singapore}
\affiliation{Centre for Quantum Technologies, National University of Singapore, Singapore 117543, Singapore}

\author{Aoqian Shi}%
\altaffiliation[]{These authors contributed equally to this work.}
\affiliation{Key Laboratory for Micro/Nano Optoelectronic Devices of Ministry of Education, School of Physics and Electronics, Hunan University, Changsha 410082, China}
\author{Jianjun Liu}%
\email{jianjun.liu@hnu.edu.cn}
\affiliation{Key Laboratory for Micro/Nano Optoelectronic Devices of Ministry of Education, School of Physics and Electronics, Hunan University, Changsha 410082, China}
\affiliation{Greater Bay Area Institute for Innovation, Hunan University, Guangzhou 511300, China}

\author{Jiangbin Gong}%
\email{phygj@nus.edu.sg}
\affiliation{Department of Physics, National University of Singapore, Singapore 117542, Singapore}
\affiliation{Centre for Quantum Technologies, National University of Singapore, Singapore 117543, Singapore}

\begin{abstract}
Criticality-based quantum sensing exploits hypersensitive response to system parameters near phase transition points. This work uncovers two metrological advantages offered by topological phase transitions when the probe is prepared as topological edge states. Firstly,
the order of topological band touching is found to determine how the metrology sensitivity scales with the system size. Engineering a topological phase transition with higher-order band touching is hence advocated, with the associated quantum Fisher information scaling as $ \mathcal{F}_Q \sim L^{2p}$, with $L$ the lattice size in one dimension, and $p$ the order of band touching. 
Secondly, with a topological lattice accommodating degenerate edge modes (such as multiple zero modes), preparing an $N$-particle entangled state at the edge and then adiabatically tuning the system to the phase transition point grows quantum entanglement to macroscopic sizes, yielding $\mathcal{F}_Q \sim N^2 L^{2p}$. Favorable scalability is still achievable through multi-criticality even when state preparation time and the shrinking sensing parameter window are taken into account. This work hence paves a possible topological phase transition-based route to harness entanglement, large lattice size, and high-order band touching for quantum metrology.
\end{abstract}

\maketitle

\section{Introduction}
As one major pillar of modern quantum technologies, quantum metrology exploits quantum features to facilitate high-resolution sensors in measuring time~\cite{ludlow2015optical}, electromagnetic fields~\cite{budker2007optical,taylor2008high, dolde2011electric,tanaka2015proposed,facon2016sensitive}, and gravitational fields~\cite{aasi2013enhanced,tino2019sage}, etc. The uncertainty of estimating an unknown parameter $\lambda$ is bounded by the Cram\'{e}r-Rao inequality~\cite{braunstein1994statistical,cramer1999mathematical}, namely $\delta\lambda^2\ge (M\mathcal{F}_{Q})^{-1}$, where $\delta\lambda^2$ is the variance, $M$ is the number of independent trials, and $\mathcal{F}_{Q}$ is Quantum Fisher Information (QFI) quantifying the amount of extractable information about $\lambda$ encoded in the state of probe, denoted $\rho_\lambda$. The quantum metrological advantage can be studied by examining how $\mathcal{F}_{Q}$ scales with physical resources, such as time, system size, or number of excitations. In particular, quantum critical systems have been viewed as promising platforms for ultra-sensitive sensors, offering advantageous scalability near phase transition points due to their divergent susceptibility to system parameters~\cite{zanardi2008quantum,invernizzi2008optimal,gammelmark2011phase,rams2018limits,chu2021dynamic,liu2021experimental,mirkhalaf2021criticality,di2023critical,MONTENEGRO20251,sarkar2025exponentially}. 

Along this avenue of criticality-based quantum metrology, the use of topological phase transitions attracted some attention~\cite{PhysRevLett.129.090503,Sarkar_2024,PhysRevLett.133.120601}. Unlike second-order phase transitions governed by local order parameters~\cite{zanardi2006ground,zanardi2007mixed,gu2008fidelity,gu2010fidelity,gammelmark2011phase,rams2018limits,montenegro2021global,he2023,adani2024critical}, the hallmark of topological phase transitions is a jump in the topological invariants characterizing symmetry-protected topological phases~\cite{RevModPhys.88.035005}. Built upon existing knowledge about topological phase transitions~\cite{PhysRevB.95.075116,PhysRevLett.120.057001,PhysRevB.107.205114,PhysRevB.111.085148,cv5q-8t25}, one major aim of this work is to identify the order of band touching as the key bulk-band feature that is directly responsible for the scaling behavior of criticality-enabled sensitivity.  
 
Such criticality-based sensitivity can be accessed not only by the bulk states, but also by topological edge states. This is because, under some adiabatic control, topological edge states can become completely delocalized states at a topological phase transition point.  The second aim of this work is to advocate the use of in-gap sharply localized topological edge states, robust against local disorder and impurities~\cite{RevModPhys.82.3045,RevModPhys.83.1057} as the initial state preparation to implement criticality-based metrology.  In particular, with multiple topological edge modes emerging at the same edge, a topological lattice can store multipartite entanglement~\cite{Wang:19,WANG2022100003,dai2022topologically,jin2025topological}, thus allowing us to harness both criticality and entanglement via adiabatic control of topological edge modes.

The main findings of this work are twofold. First, the QFI divergence behavior at criticality is found to be determined by the order of band touchings, with its scaling given by $L^{2p}$,  where $L$ is the lattice size and $p$ is the order of band touching, achievable by position measurements. Furthermore, multi-criticality is found to be a delicate resource to design specific adiabatic paths towards criticality. Second, exploiting degenerate topological zero modes in a class of topological lattice systems, preparing an $N$-particle entangled state at one edge and leveraging an adiabatic encoding scheme, the overall QFI is found to scale as $N^2 L^{2p}$, thus revealing an unprecedented possibility in the use of entanglement, large system size, and high-order band touching all together to boost quantum metrology.

\section{Quantum parameter estimation preliminaries}
In a general quantum parameter estimation protocol, a key figure of merit is the amount of maximally extractable information carried by a probe. First, we prepare an initial state $\rho_0$ and then encode an unknown parameter $\lambda$ via dynamical evolution, such as quench or adiabatic ramp, thereby producing a parameterized state $\rho_{\lambda}$. The QFI, quantifying the information about $\lambda$ encoded in the probe, is denoted $\mathcal{F}_{Q}$ below and can be found from~\cite{gu2007}
\begin{equation}
    \mathcal{F}_{Q}=8\lim_{\delta\lambda\to 0} \frac{1-F(\lambda,\lambda+\delta\lambda)}{\delta\lambda^2},
\label{Eq:derivative_QFI}
\end{equation}
where $F(\lambda,\lambda+\delta\lambda)$ is the fidelity between two neighboring states in the parameter manifold, namely $F(\lambda,\lambda+\delta\lambda)
= \text{Tr}^2
   \sqrt{\,
      \rho_{\lambda}^{1/2}\,
      \rho_{\lambda+\delta\lambda}\,
      \rho_{\lambda}^{1/2}\,
   }$, with $\delta\lambda$ being an infinitesimal differentiation with respect to the parameter $\lambda$. 

We focus on a metrology scenario where the state is initially prepared as one eigenstate (indexed by $l_0$) of the topological lattice Hamiltonian (e.g., one bulk state or one topological edge state).  Adopting the  non-degenerate perturbation theory~\cite{gu2007} (analogous to the treatment of the quantum geometric tensor~\cite{provost1980riemannian,zhang2019}), one finds
\begin{equation}
    \mathcal{F}_{Q}=4\sum_{l\neq l_0}\frac{|\langle E_l|H_r|E_{l_0}\rangle|^2}{|E_l-E_{l_0}|^2},
\label{Eq:perturb_QFI}
\end{equation}
where $E_l$ and $|E_l\rangle$ are the eigenvalue and corresponding eigenvector indexed by $l$, and $H_r$ is the driving Hamiltonian coupling linearly with the unknown parameter $\lambda$. If there are degeneracies (not elaborated below), the QFI calculations can be connected with the non-Abelian quantum geometric tensor~\cite{ma2010,ding2022,PhysRevA.108.032218,ding2024}. Because the energy gap between state $E_{l_0}$ and its nearest state contributes most significantly to the QFI, it is lower bounded by the inverse of the square of the minimal energy gap $(E_{l}-E_{l_0})$, which follows a power-law function with system size $L$ as $(E_{l}-E_{l_0}) \sim L^{-z}$ with $z$ being the dynamical critical exponent~\cite{hohenberg1977,sondhi1997,Sachdev_2011}.
The numerator in Eq.~(\ref{Eq:perturb_QFI}) introduces additional scaling with system size or particle number depending on the nature of $H_r$~\cite{he2023,Yousefjani_2023,he2025}. Combining these observations yields the operator-norm bound of the QFI: $\mathcal{F}_{Q}\le ||H_r||^2 \Delta E^{-2}$, analytically shown in Ref.~\cite{Abiuso2025}.

\begin{figure*}
  \includegraphics[scale=0.23]{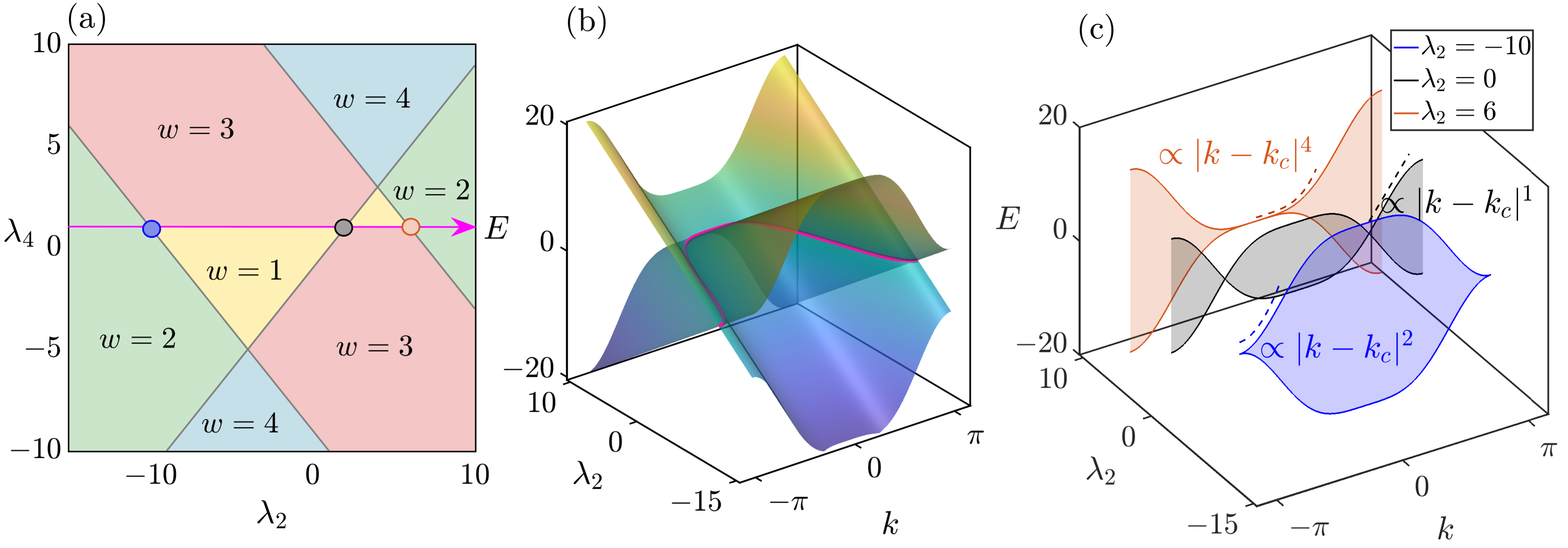}
  \caption{(a) Phase diagram characterized by winding number $w$ of the eSSH model with intercell long-range coupling range up to $R=4$. The pink arrow shows the parameter-sweeping range used to evaluate the band structure, while fixing $\lambda_4=1$ in panel (b). The blue, black, and orange points denote the band-touching curves with regard to the corresponding colors. The rest parameters are set as $(\lambda_0,\lambda_1,\lambda_3)=(1,-4,-4)$. (b) Band structure of our model as a function of momenta $k$ and next-nearest-neighbor coupling strength $\lambda_2$, where the pink solid line highlights the band touching condition $E(\lambda_2,k)=0$. (c) By choosing $\lambda_2=-10,0,6$, various orders of band-touching near the critical momentum $k_c$ are depicted in blue, black, and orange colors, where the algebraic convergence law of energy gap fits $E\propto|k-k_c|^{p}$ with $p=2,1,4$ respectively.}
  \label{fig:Fig1}
\end{figure*}

\section{On the order of band touching}
Important features of a topological phase transition include the change in the topological invariant before and after band touching, the presence of multi-critical points, the number of band touching points, and the order of band touching.  As seen below, the order of band touching is found to solely determine the scaling of the QFI at criticality.  Let the direct energy gap $\Delta E$ of two nearby bulk states at the same Bloch momentum $k$ follow a power-law algebraic function near a critical momentum, namely,  $\Delta E\sim|k-k_c|^p$, with $p$ being the order of band touching. In particular, $p=1,2,3,4,\ldots$ correspond to linear, quadratic, cubic, quartic, and even higher-order band touching, triggering a topological phase transition. Without the loss of generality, we assume that the topological band is symmetric and centered at $E=0$, with a positive branch described by an analytical function $E(\lambda,k)\ge0$. A multi-variable Taylor expansion about the critical point $(\lambda_c,k_c)$ yields
\begin{equation}
    E(\lambda_c+\delta\lambda,k_c+\delta k)
    =\sum_{n_1,n_2=0}^{\infty}\frac{1}{n_1!n_2!}[\partial_{\lambda}^{n_1}\partial_k^{n_2}  E(\lambda_c,k_c)](\delta\lambda)^{n_1}(\delta k)^{n_2}.
\label{Eq:Analytic_func}
\end{equation}
By designing system parameters so that all terms as a power function of $\delta k$ below the power index $p$ vanish in the expansion above, we obtain a pure $p$-th-order band touching. To the leading order of $\delta\lambda$, one finds
\begin{equation}
    \Delta E(\lambda)\approx a\delta\lambda+b_p(\delta k)^p+O[(\delta\lambda)^2,\delta\lambda(\delta k)^{p},(\delta k)^{p+1}],
\label{Eq:Delta E_expansion}
\end{equation}
where $\delta k=|k-k_c|$ and $\delta\lambda=|\lambda-\lambda_c|$, and $a$ and $b_p$ are prefactors.

To be specific, consider now a one-dimensional lattice with size $L$ under the periodic boundary condition (analogous arguments can be made for the open boundary condition). Because the Bloch states possess quantized momenta $k_n=2\pi n/L$, $\delta k$ in Eq.~(\ref{Eq:Delta E_expansion}) as the nonzero momentum deviation from $k_c$ is given by $\delta k\equiv |k_n-k_{n-1}|=2\pi/L\sim L^{-1}$.
Substituting this into Eq.~(\ref{Eq:Delta E_expansion}) yields the power-law scaling of $\Delta E$ at $\lambda=\lambda_c$, namely, $\Delta E\sim (\delta k)^p \sim L^{-p}$. That is, the energy gap at $\lambda=\lambda_c$ in the vicinity of the band touching point $k_c$ scales as $\Delta E \sim L^{-p}$.  Connecting this result with the above minimal energy gap scaling relation  $(E_l-E_{l_0})\sim L^{-z}$, one immediately sees that the dynamical exponent $z$ is given by $z=p$. As such, the QFI scaling dictated by a band touching of order $p$ is given by 
\begin{equation}
    \mathcal{F}_{Q} \sim ||H_r||^2 \Delta E^{-2} \sim ||H_r||^2 L^{2p}.
\label{Eq:QFI_scaling}
\end{equation}

\section{One-dimensional lattice model with long-range hopping}
As an explicit topological lattice model, we consider an extended Su–Schrieffer–Heeger (eSSH) model depicting a one-dimensional dimerized chain of $L$ unit cells with sublattices $A$ and $B$ per cell~\cite{PhysRevB.103.224208,PhysRevApplied.19.054028,PhysRevResearch.6.043087}, with its Hamiltonian given by
\begin{equation}
    H=\sum_{r=0}^R \lambda_rH_r=\sum_{r=0}^{R}\sum_{j=1}^{L-r}\lambda_r a_{j+r}^{\dagger}b_j+h.c.,
\label{Eq:Hamiltonian_real_space}
\end{equation}
where $a_j^{\dagger}(a_j)$, $b_j^{\dagger}(b_j)$ are the creation (annihilation) operator at sublattice $A$ and $B$ in $j$-th unit cells. $\lambda_0$ is the intracell hopping strength while $\lambda_{r\neq0}$ denotes the intercell hopping strength between $j$-th unit cell and its $r$-th neighboring unit cells. Later, we shall view one of the parameters $\lambda_{r}$, e.g., $\lambda_2$ as the above-defined parameter $\lambda$ to be estimated.   In the momentum space, the Bloch Hamiltonian assumes the following chiral form
\begin{equation}
    H(\lambda,k)=\begin{pmatrix}
        0 & h(\lambda,k) \\
        h^*(\lambda,k) & 0
    \end{pmatrix}
\label{Eq:Hamiltonian_momentum_space}
\end{equation}
with
\begin{equation}
    h(\lambda,k)=\sum_{r=0}^R \lambda_r e^{ikr},
\end{equation}
where quasi-momentum $k$ lies within the first Brillouin zone. The Bloch Hamiltonian in Eq.~(\ref{Eq:Hamiltonian_momentum_space}) preserves sublattice (chiral) symmetry, namely $\sigma_zH(k)\sigma_z^{\dagger}=-H(k)$, which enforces spectral symmetry about zero energy. The topological phases are characterized by the following winding number $w$~\cite{PhysRevB.103.224208,PhysRevApplied.19.054028,PhysRevResearch.6.043087}, 
\begin{equation}
    w(\lambda)=\frac{1}{2\pi i}\int_{-\pi}^{\pi}\!dk\;\partial_k \text{ln}h(\lambda,k).
\label{Eq:winding_number}
\end{equation}

To access richer topological phases and 
high-order band touchings, we set the hopping range up to $R=4$.  As one example for illustration, we first fix  $(\lambda_{0},\lambda_{1},\lambda_{3})=(1,-4,-4)$ and then investigate the phase diagram in the parameter space $(\lambda_2, \lambda_4)$, with multiple phase boundaries and multi-critical points, as shown in Fig.~\ref{fig:Fig1}(a).  The chosen $(\lambda_{0},\lambda_{1},\lambda_{3})$ can reduce the polynomial spectral function $h(\lambda,k)=\sum_r\lambda_{r}e^{ikr}$ to $h(\lambda,k)=(e^{ik}-e^{ik_c})^p=0$ at some critical momentum $k_c$, thus yielding a $p$-th-order band touching.  A variety of topological phases with winding number ranging from 1 to 4 are depicted in Fig.~\ref{fig:Fig1}(a). To highlight the band-touching behavior along a concrete path, we fix $\lambda_4=1$ and sweep $\lambda_2$ along the trajectory indicated by a pink arrow in Fig.~\ref{fig:Fig1}(a). In Fig.~\ref{fig:Fig1}(b), we plot the dispersion $E(\lambda_2,k)$ versus $\lambda_2$ and $k$. The two bands close and reopen across the first Brillouin zone as we vary $\lambda_2$, where the band-touching condition, $E(\lambda_2,k)=0$, is tracked by a pink solid line. In Fig.~\ref{fig:Fig1}(c), we show the band structure at three representative long-range tunneling strengths, namely $\lambda_2=-10,0,6$, with algebraic band closings as $E\propto |k-k_c|^{p}$ with $p=2,1,4$, corresponding to the blue, black, and orange dots in Fig.~\ref{fig:Fig1}(a), respectively.

\begin{figure*}
  \includegraphics[scale=0.23]{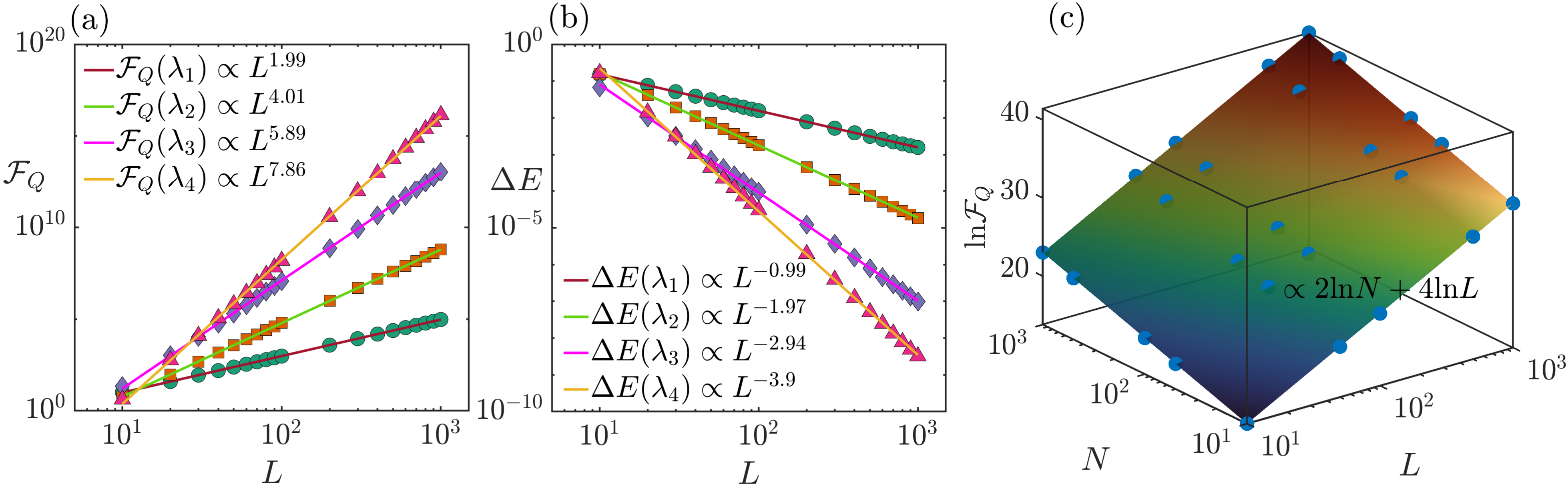}
  \caption{(a) The QFI and (b) the bulk-edge energy gap $\Delta E$ vs lattice size $L$ at different topological phase transition points with different orders of band touching, with the hopping range from $R=1$ at $(\lambda_{0},\lambda_{1})=(1,-1)$ to $R=4$ at $(\lambda_{0},\lambda_{1}, \lambda_{2},\lambda_{3},\lambda_{4})=(1,-4,6,-4,1)$. Symbols denote the $\mathcal{F}_{Q}$ and $\Delta E$, whereas solid lines indicate the power-law fitting curves, verifying the algebraic relation between scaling exponent and band-touching order as $\beta\simeq 2p$. (c) The QFI of the non-interacting many-body probe as a function of both system size $L$ and particle number $N$ with the range of intercell hopping $R=2$ at the phase transition point $(\lambda_{0}, \lambda_{1},\lambda_{2})=(1,2,1)$. Symbols are for computational results, whereas the surface plot denotes the power fitting function $\mathcal{F}_Q\propto N^2L^{2p}$ with $p=2$.}
  \label{fig:Fig2}
\end{figure*}

To verify the above-derived QFI scaling behavior at topological criticality, we numerically investigate the QFI as well as the bulk-edge energy gap vs the system size $L$, at several chosen topological transition points. To that end, we also consider examples with different hopping ranges $R$.   In Figs.~\ref{fig:Fig2}(a) and \ref{fig:Fig2}(b), we show the results for $p=1,2,3,4$. No matter which $\lambda_r$ is viewed as the unknown parameter $\lambda$, the associated driving Hamiltonian $H_r$ here has a size-independent spectral semi-norm, namely $||H_r||\sim L^{\alpha}$ with $\alpha=0$.  As our computational results show, both the QFI and the gap exhibit power-law scaling, $\mathcal{F}_{Q}\sim L^{\beta}$ and $\Delta E\sim L^{-p}$ with exponents obeying $\beta=2(p+\alpha)$, in agreement with the scaling analysis above.

The scaling behavior of the computed QFI can be qualitatively connected with the order of band touching as follows.  Low-energy physics is governed by the Bloch momentum states near the band touching point $k_c$ at the topological transition point $\lambda=\lambda_c$. If $p$ is to increase, the energy dispersion flattens, enhancing the local density of states near $k_c$. Consequently, more low-energy modes can be excited near $k_c$, thus facilitating an amplified parametric response to a small shift in $\lambda$ from $\lambda_c$.
Because this connection between QFI scaling and the order of band touching is previously unknown~\cite{PhysRevLett.129.090503},
in the Appendices~\ref{App:eSSH_derive} and \ref{App:2D},  we provide a proof for such a connection via an asymptotic approach, and also extend our analysis to higher-order topological insulators~\cite{PhysRevLett.128.127601,PhysRevLett.131.157201} and Chern insulators~\cite{Bernevig+2013,RevModPhys.95.011002}. These additional results further confirm that our finding also applies to other topological symmetry classes.

\section{Optimal measurement}
The metrological advantage is meaningful only if it can be extracted with a practical measurement. Such precision is evaluated through a basis-dependent quantity, named classical Fisher information~\cite{Matteo} with $\mathcal{F}_C\le \mathcal{F}_Q$, saturable through the optimal measurement basis maximizing $\mathcal{F}_C$. In the Appendix~\ref{App:Opt_Measure}, we show that the optimal readout is naturally provided by the position basis since there is no need to capture the site-dependent relative phases of the zero-energy edge mode at the criticality. Therefore, position measurements achieve the same critical scaling of $\mathcal{F}_C$ as $\mathcal{F}_Q$, showing that the enhanced sensitivity is experimentally feasible in principle.

\section{Edge entanglement storage as a probe}
It is now well understood from previous work~\cite{gio2004,gio2006,giovannetti2011advances} that entanglement, such as that in GHZ- and N00N-like states, can be used to achieve the Heisenberg scaling of QFI with the number of particles $N$, i.e. $\mathcal{F}_Q\sim N^2$. One important question is whether the above criticality-based metrology can be further enhanced by loading more particles to the lattice. With the initial probe state prepared as entangled topological edge modes, we show below, within an ideal noninteracting model naturally realizable in photonic platforms where spatial mode overlap does not by itself induce interactions, that achieving Heisenberg scaling with $N$ is feasible.

For system parameters away from criticality, topological edge modes are sharply localized and well protected by lattice symmetry. With proper topological lattice design, a topological lattice can store entanglement via its edge modes~\cite{Wang:19,WANG2022100003,dai2022topologically,jin2025topological}.    In the eSSH lattice with chiral symmetry elaborated above, two or even more edge states are localized at the same edge, all pinned at zero energy. To push for the best metrological performance of a topological lattice, we now consider utilizing entanglement between different edge modes,  pre-stored within the in-gap zero-energy degenerate subspace. 

We now consider the following GHZ state as the probe, with the two components being the two left-localized zero-energy edge modes at $\lambda=\lambda(0)$ away from $\lambda_c$, i.e. $|\Phi_0\rangle=(|L_1\rangle^{\otimes N}+|L_2\rangle^{\otimes N})/\sqrt{2}$.  This is then followed by an adiabatic control protocol, under the slowly time-dependent Hamiltonian $H({\lambda}(t))=\lambda(t)H_2+\sum_{r\neq 2}\lambda_rH_r$ driven towards the topological phase transition point $\lambda_c=\lambda(T)$, with $T$ the total duration of the adiabatic protocol.
Because this probe state is in the zero-energy subspace, the exact start time of the protocol does not lead to any consequence (hence stable entanglement storage).  The final parameterized state reads
\begin{align}
    |\Phi_{\lambda}(T)\rangle&=\frac{[U_{\lambda}(T,0)|L_1\rangle]^{\otimes N}+[U_{\lambda}(T,0)|L_2\rangle]^{\otimes N}}{\sqrt{2}},
\end{align}
where $U_\lambda(T,0)$ is the time evolution operator associated with the adiabatic protocol, with $U_{\lambda}(T,0)=\mathcal{T}\text{e}^{-i\int_0^{T}H(\tau)d\tau}$,  $\mathcal{T}$ the time-ordering operator. $T$ should scale as $T\gg\Delta E^{-1}\sim L^z$~\cite{hohenberg1977},  stipulated by the critical slowing down, with $z=p$. Indeed, the energy gap between the bulk and the edge states at $\lambda_c$ scales as $L^{-p}$.
The QFI with respect to $\lambda$ is then evaluated through $\mathcal{F}_Q=4[\langle\partial_{\lambda}\Phi_{\lambda}|\partial_{\lambda}\Phi_{\lambda}\rangle-|\langle \Phi_{\lambda}|\partial_{\lambda}\Phi_{\lambda}\rangle|^2]$~\cite{Matteo}. Under adiabatic conditions, $U_{\lambda}(T,0)$ can be assumed to be unitary with respect to the subspace accommodating all zero-energy edge modes. 


To proceed, we define two single-particle branches
\begin{equation}
\ket{\psi_\mu}\equiv U_\lambda(T,0)\ket{L_\mu},\quad
\ket{\psi'_\mu}\equiv \partial_\lambda\ket{\psi_\mu}=[\partial_\lambda U_\lambda(T,0)]\ket{L_\mu},
\end{equation}
where $\mu = 1,2$.
Some straightforward calculations take us to the following  compact 
\begin{equation}
\braket{\partial_\lambda(\psi_\mu^{\otimes N})|{\partial_\lambda(\psi_\mu^{\otimes N})}}
= N(N-1)|a_\mu|^2+N\,b_\mu,
\end{equation}
where
\begin{align}
a_\mu&\coloneqq \braket{\psi_\mu|{\psi'_\mu}}
=\bra{L_\mu}[U_\lambda^\dagger(T,0)\partial_\lambda U_\lambda(T,0)]\ket{L_\mu},\\
b_\mu&\coloneqq \braket{\psi'_\mu|{\psi'_\mu}}
=\bra{L_\mu}[\partial_\lambda U_\lambda^\dagger(T,0)][\partial_\lambda U_\lambda(T,0)]\ket{L_\mu}.
\end{align}
Therefore,
\begin{equation}
\braket{\partial_\lambda\Phi_\lambda|{\partial_\lambda\Phi_\lambda}}
=\frac12\Big[N(N-1)(|a_1|^2+|a_2|^2)+N(b_1+b_2)\Big].
\end{equation}
Likewise, we obtain
\begin{align}
\braket{\Phi_\lambda|\partial_\lambda\Phi_\lambda}
&=\frac12\sum_{\mu=1,2}\braket{\psi_\mu^{\otimes N}|\partial_\lambda(\psi_\mu)^{\otimes N}}
=\frac{N}{2}\,(a_1+a_2).
\end{align}
Finally, we obtain the QFI
\begin{align}
\mathcal{F}_Q
&=4\!\left[\braket{\partial_\lambda\Phi_\lambda|\partial_\lambda\Phi_\lambda}
-|\braket{\Phi_\lambda|\partial_\lambda\Phi_\lambda}|^2\right]\notag\\
&=N^2|a_1-a_2|^2
+2N\Big[(b_1-|a_1|^2)+(b_2-|a_2|^2)\Big].
\label{Eq:FQeq}
\end{align}

To digest the above expression of QFI, 
we use $i\partial_t U_\lambda(t,0)=H_\lambda(t)U_\lambda(t,0)$ and define $A_\lambda(t)\equiv U_\lambda^\dagger(t,0)\partial_\lambda U_\lambda(t,0)$. One then finds
$\partial_t A_\lambda(t)=-i\,U_\lambda^\dagger(t,0)[\partial_\lambda H_\lambda(t)]U_\lambda(t,0)$ with $A_\lambda(0)=0$, and hence the following equation,
\begin{equation}
U_\lambda^\dagger(T,0)\partial_\lambda U_\lambda(T,0)
=-i\int_0^T\!d\tau\;U_\lambda^\dagger(\tau,0)[\partial_\lambda H_\lambda (\tau)]U_\lambda(\tau,0).
\label{Eq:UdU_exact_compact}
\end{equation}
In our adiabatic protocol for the eSSH model, $\partial_\lambda H_\lambda(\tau)=H_2$, and hence,
\begin{equation}
a_\mu=-i\int_0^T\!d\tau\;\langle L_\mu|U_\lambda^\dagger(\tau,0)H_2 U_{\lambda}(\tau,0)|L_\mu\rangle.
\label{Eq:a}
\end{equation}
Eq.~(\ref{Eq:FQeq}) indicates that it is necessary and sufficient to have $(a_1-a_2)\ne 0$ to obtain Heisenberg scaling with $N$. This being the case and noting that $\|H_2\|\sim O(1)$, the integrand is $O(1)$ and thus $a_\mu\sim T\sim L^p$ due to the critical slowing down, implying $|a_1-a_2|^2\sim L^{2p}$. The difference $(a_1-a_2)$ can be interpreted as the mismatch of the time-accumulated expectation value $\langle H_2\rangle$ along the two edge-mode branches comprising the entangled state prepared as the probe.

We now return to the above eSSH model with band touching order $p=2$ and adiabatically drive the initial GHZ state from $(\lambda_{0}=1,\lambda_{1}=2,\lambda_2>1)$ to the multi-critical point $(\lambda_{0}=1,\lambda_{1}=2,\lambda_{2}=1)$ around which the winding number jumps from 2 to 1~\cite{PhysRevResearch.6.043087}. Notably, although the energy gap within the edge-mode degenerate subspace is exponentially small as the system size $L$ increases, there is no rotation between the two degenerate modes undergoing adiabatic following along the adiabatic path because of chiral symmetry protection. Specifically, before reaching the critical point $\lambda_c$, the adiabatic protocol does not incur any transitions between the two left-edge localized edge modes -- the driving Hamiltonian  $H_2$ only has hopping between two different sublattices, but both edge modes only occupy the same sublattice $A$, yielding a vanishing matrix element for $H_2$ between the two left-edge modes.
Furthermore, as the winding number decreases from 2 to 1, at the topological phase transition point, one of the two edge modes will become a bulk state, and the other edge mode becomes delocalized as well but still occupies the sublattice $A$ only.  As such, $a_1$ and $a_2$, the two time-integrated expectation values of $H_2$ over the two evolving trajectories $U_{\lambda}(t,0)|L_\mu\rangle$, must be different, thus ensuring the above-mentioned Heisenberg scaling with $N$.
Computationally, we have verified the adiabatic following of the two edge-mode branches from numerical results shown in Appendix~\ref{App:Multi-particle} and that the total evolution time $T\sim 1/\Delta$ is enough for us to see the asymptotic scaling of the QFI.
Under the adiabatic following assumption, the above expression for the QFI  can be further reduced. A detailed derivation of the QFI under the adiabatic following assumption is presented in Appendix~\ref{App:Multi-particle}. The QFI near $\lambda_c$ as a function of both system size $L$ and particle number $N$ can then be numerically obtained and shown in Fig.~\ref{fig:Fig2}(c). There, the QFI is seen to follow precisely our prediction above, namely,  $\mathcal{F}_Q\sim N^2L^{2p}$ with $p=2$.

\section{Multi-criticality as a resource}
While the order of band touching $p$ characterizes favorable probe scalability, the critical window within which the asymptotic scaling of QFI is available, defined by the correlation length $\xi\sim L\sim w_c^{-\nu}=|\lambda-\lambda_c|^{-\nu}$, is equally significant. To this end, we define a new figure of merit to account for the shrinking size of the sensing window, denoted $w_c$, as a penalty.  This takes us to the window-regularized QFI (WQFI) $\mathcal{W}\equiv w_c^2\mathcal{F}_Q$. Such a quantity can be interpreted straightforwardly as the infidelity between two nearby states in the parameter space, up to the second-order expansion in $\delta\lambda$. Evaluating WQFI over $\delta\lambda\lesssim w_c$ shows that it quantifies the total distinguishability accumulated across the accessible critical window. At first glance, WQFI exhibits unfavorable scaling $\mathcal{W}\sim N^2L^{2p-\frac{2}{\nu}}\sim N^2L^0$ since generally $\nu p=\nu z=1$ for topological phase transitions to occur. However, one can always design specific paths that approach the same multi-critical point, along which a larger $\nu$ is available regardless of the band-touching order, thereby recovering favorable scaling behavior of WQFI still dependent on $p$ (see the detailed proof in Appendix~\ref{App:Multicriticality}).  Indeed, even after taking state preparation time into account~\cite{rams2018limits,he2023}, the band-touching order $p$ itself remains significant for the scaling of WQFI, as manifested in the scaling $\mathcal{W}/T\sim N^2L^{p-\frac{2}{\nu}}$.


\section{Conclusion}
The use of topological lattices for quantum metrology was also previously considered in connection with some remarkable non-Hermitian physics arising from nonreciprocal couplings or gain/loss~\cite{PhysRevLett.125.180403, mcdonald2020exponentially,PhysRevResearch.4.013113, deng2024ultrasensitive,parto2025enhanced,PhysRevResearch.7.013309,8zfx-c7mw,f6wd-gljq}. In these scenarios, it is hard to harness entanglement for quantum sensing advantages because the non-Hermitian setting is not easily compatible with the use of quantum entanglement. This work uncovers how entanglement stored in topological edge modes can be an important resource for criticality-based quantum metrology.  To that end, we also find a direct connection between the order of band touching and the scaling of the probe sensitivity with the lattice size: the scaling exponent of QFI increases as the order of the band touching goes higher. It is hence worthwhile searching for lattice designs that can yield a band touching of two flat bands, namely, a band touching at the infinite order and hence the best scalability of the QFI. We have shown that position measurements suffice to achieve the theoretical precision. Although the scalability of the QFI with the system size may vanish when considering both the cost of state preparation time and the shrinking window for criticality-based sensing, we have shown that taking specific paths in the parameter space, leveraging on multi-criticality, can recover useful sensitivity scaling behavior. It should also be acknowledged that the order of band touching, the universality class~\cite{PhysRevB.106.134203}, and the associated topological edge modes discussed in this work are expected to be robust against perturbations provided that the chiral symmetry is preserved (as verified in various platforms~\cite{PhysRevApplied.19.054028,PhysRevResearch.6.043087}). In future work, we aim to extend our study to many-body lattice modes and also complement our study by developing protocols that actively generate edge-mode entanglement.  To fight against decoherence effects and hence partially retain the metrological advantage of quantum entanglement in realistic noisy situations~\cite{PhysRevLett.131.050801,lkrt-lvng},
the key idea of this work can be extended to non-equilibrium Floquet topological lattices with multiple degenerate edge modes~\cite{PhysRevB.87.201109,PhysRevB.90.195419,PhysRevB.99.045441}.

\section*{Acknowledgments}
We acknowledge Zhixing Zou and Fei Yu for fruitful discussions. J.G. acknowledges support by the National Research Foundation, Singapore, through the National Quantum Office, hosted in A*STAR, under its Centre for Quantum Technologies Funding Initiative (S24Q2d0009). J.L. is supported by the National Natural Science Foundation of China (Grant No. 62575099). A.S. acknowledges support by the China Scholarship Council.

\section*{Data availability}
There is no publicly disclosed research data supporting this article. Further data and information are available upon reasonable request.

\begin{widetext}
\onecolumngrid
\appendix
\renewcommand{\appendixname}{APPENDIX}

\appsection{\MakeUppercase{An asymptotic solution to QFI scalability in eSSH model}}\label{App:eSSH_derive}
Following the main text, we give a step-by-step proof of the scaling relation by analyzing the asymptotic form of the zero-energy edge mode near the topological phase transition. We consider the extended Su-Schrieffer-Heeger (eSSH) model with open boundary conditions and $L$ unit cells (two sublattices $A,B$ per cell),
\begin{equation}
    H=\sum_{r=0}^{R}\sum_{j=1}^{L-r}\lambda_r a_{j+r}^{\dagger}b_j+h.c.,
\label{Eq:SM_Hamiltonian_real_space}
\end{equation}
where $a_j^{\dagger}(a_j)$, $b_j^{\dagger}(b_j)$ are creation (annihilation) operator at sublattice $A$ and $B$ in $j$-th unit cells. $\lambda_0$ is the intracell hopping strength while $\lambda_{r\neq 0}$ denotes the intercell hopping strength between $j$-th unit cell and its $r$-th neighboring unit cells. Now we consider the static eigenfunction of the zero-energy edge state localized on the left boundary in cell $A$, which can be written as
\begin{equation}
    |\psi_L\rangle=\sum_{j=1}^L \phi_j a_j^{\dagger}|0\rangle,
\label{Eq:edge_state_ansatz}
\end{equation}
where $\phi_j$ denotes the probability amplitude of the state at the sublattice $A$ in cell $j$. By substituting the eigenfunction into Eq.~(\ref{Eq:SM_Hamiltonian_real_space}), we have
\begin{align}
        H|\psi_L\rangle =0\quad &\Rightarrow \sum_{j=1}^{L-r}\sum_{r=0}^R (\lambda_r a_{j+r}^{\dagger}b_j+\lambda_r^{*}b_j^{\dagger}a_{j+r})\sum_{j'=1}^L \phi_{j'}a_{j'}^{\dagger}=0, \nonumber\\
        & \Rightarrow \sum_{j=1}^{L-r}\sum_{j'=1}^L\sum_{r=0}^R (\lambda_r\phi_{j'}a_{j+r}^{\dagger}b_ja_{j'}^{\dagger}|0\rangle+\phi_{j'}\lambda_r^{*}b_j^{\dagger}a_{j+r}a_{j'}^{\dagger}|0\rangle)=0, \nonumber\\
        & \Rightarrow \sum_{j=1}^{L-r}\sum_{j'=1}^L\sum_{r=0}^R [\lambda_r\phi_{j'}a_{j+r}^{\dagger}a_{j'}^{\dagger}b_j|0\rangle+\phi_{j'}\lambda_r^{*}b_j^{\dagger}(\delta_{j',j+r}a_{j'}^{\dagger}a_{j+r})|0\rangle]=0, \nonumber\\
        & \Rightarrow \sum_{j=1}^{L-r}\sum_{j'=1}^L \sum_{r=0}^R \lambda_r^{*}\phi_{j'}\delta_{j',j+r}b_j^{\dagger}|0\rangle=0, \nonumber\\
        & \Rightarrow \sum_{j=1}^{L-r} \sum_{r=0}^R \lambda_r^{*}\phi_{j+r}b_j^{\dagger}=0.
\end{align}
By projecting both sides onto the basis state $b_j^{\dagger}|0\rangle$, we eliminate the sum over cell sites and end up with the recursion relation between probability amplitude $\phi_{j+r}$ and parameter $\lambda_r^*$
\begin{equation}
    \sum_{r=0}^R \lambda_r^*\phi_{j+r}=0.
\label{Eq:recursion_relation}
\end{equation}

Not bothering to write the conjugate sign in the following derivation, we denote $\Lambda_r\equiv\lambda_r^*$ and $\Lambda_r\in \mathbb{C}$. We start with a polynomial trial $\phi_j=z^j$ with $z$ being a complex number. Substituting the trial solution back into Eq.~(\ref{Eq:recursion_relation}) yields a polynomial equation
\begin{equation}
    P(z,\Lambda)=\sum_{r=0}^R \Lambda_r z^r=0.
\end{equation}
Now we consider that the band-touching order is purely of $p$, and the system is exactly at the phase transition point $\Lambda=\Lambda_c$.
There exist roots falling onto a unit circle $|z_*|=1$ with $p$-fold degeneracy
\begin{equation}
    P(z)=f(z)\cdot(z-z_*)^p,\quad  f(z_*)\neq 0.
\end{equation}
The probability amplitude derived from the solution to $P(z)=0$ is nothing but a linear combination of polynomials, namely $\phi_j^*=z_*^j\sum_{s=0}^{p-1}c_sj^s$ with $c_s$ the coefficient. The eigenstate with zero energy is dominated by an algebraic envelop as $\phi_j^*\propto j^{p-1}$.

When the system parameter deviates from the criticality by an infinitesimal amount of $\delta u$, the polynomial becomes
\begin{equation}
    P(z,\Lambda)=f(z)\cdot(z-z_*)^p+g(\Lambda)\cdot \delta \Lambda+O\!\left[(z-z_*)^{p+1},\delta \Lambda(z-z_*)\right].
\end{equation}
The solution to $P(z,\Lambda)=0$ is no longer the roots with $p$ multiplicity but split into $p$ near-degenerate roots, namely
\begin{equation}
    z_s-z_*=\left[-\frac{g(\Lambda)}{f(z_j)}\delta \Lambda\right]^{\frac{1}{p}}=d_s(\delta \Lambda)^{\frac{1}{p}},\quad s=1,2,\ldots,p,
\label{Eq:split_solution}
\end{equation}
with $d_s$ a constant. Here, we assume $\phi_j=q_s^j\phi_j^*\propto (q_sz_*)^jj^{p-1}$, with a complex number $|q_s|\approx 1$, and $z_s=q_sz_*$. Substituting $z_s=q_sz_*$ into Eq.~(\ref{Eq:split_solution}) yields
\begin{equation}
    (q_s-1)^p=d_s(\delta \Lambda).
\end{equation}
Thus, $q_s$ itself also resides in the vicinity of 1, namely $q_s=1+\delta_s$ with $\delta_s$ an infinitesimal complex number. We rewrite $q_s$ in an exponential form as
\begin{align}
    q_s&=1+\delta_s=e^{\text{ln}(1+\delta_s)}=e^{\eta_s},\quad\eta_s=-\kappa_s+i\theta_s,\quad\kappa_s,\theta_s\ll1, \nonumber\\
    &\Rightarrow z_s=q_sz_*=z_*e^{-\kappa_s+i\theta_s},\nonumber\\
    &\Rightarrow |z_s|=e^{-\kappa_s}.
\end{align}
Notably, the amplitude $\phi_j$ should be chosen so that the $|\psi_L\rangle$ is left-localized and normalizable, indicating $|z_s|<|z_*|=1$, which, in turns, implies $|\phi_j|^2\sim j^{2p-2}|q_s|^{2j}$ to decay exponentially as $j\rightarrow\infty$ from the left boundary, namely $\phi_j\sim j^{p-1}q_s^j\sim j^{p-1}e^{-\kappa_s j}e^{i\theta_s j}$. Let the localization length be $\xi\equiv1/\kappa$, we have the asymptotic scaling relation between localization length and parameter deviation $\delta \Lambda$ as
\begin{align}
    q_s-1&=e^{-\kappa_s+i\theta_s}-1\sim -\kappa_s+i\theta_s\sim C(\delta \Lambda)^{\frac{1}{p}},\nonumber\\
    \quad &\kappa_s,\theta_s\sim C(\delta \Lambda)^{\frac{1}{p}},\nonumber\\
    \quad &\xi=\frac{1}{\kappa_s}\sim C(\delta \Lambda)^{-\frac{1}{p}}.
    \label{Eq:localization_length_dlambda}
\end{align}
Comparing Eq~(\ref{Eq:localization_length_dlambda}) with the standard finite-size divergence of localization length $\xi\sim|\Lambda-\Lambda_c|^{-\nu}$ yields $\nu p=1$.

Consider evaluating the QFI with a pure state~\cite{Matteo}, the formula reads
\begin{equation}
    \mathcal{F}_{Q}=4(\langle\partial_{\Lambda}\psi|\partial_{\Lambda}\psi\rangle-|\langle\psi|\partial_{\Lambda}\psi\rangle|^2).
\label{Eq:QFI_pure_state}
\end{equation}
In general, the left-localized edge state is parameterized as $|\psi_L(\Lambda)\rangle=\sum_j\phi_j(\Lambda)a_j^{\dagger}|0\rangle$ where $\phi_j(\Lambda)=D(\Lambda)j^{p-1}e^{ik(\Lambda)j}e^{-j/\xi(\Lambda)}$, and $D(u)$ is a constant. Additionally, we define the probability distribution of the state $|\psi_L(\Lambda)\rangle$ at each unit cell as $p_j\equiv |\phi_j|^2$. For simplicity, we do not denote the explicit dependence on $u$ in the following. The derivative of probability amplitude with respect to $\Lambda$ is
\begin{equation}
    \partial_{\Lambda}\phi_j=\phi_j(\partial_{\Lambda}\text{ln}D+\frac{j}{\xi^2}\partial_{\Lambda}\xi+ij\partial_{\Lambda}k).
\end{equation}
Since the probability distribution is normalized, i.e. $\sum_jp_j=1$, we have 
\begin{align}
    &\partial_{\Lambda}\sum_jp_j=2\sum_jp_j\Re(\partial_{\Lambda}\text{ln}\phi_j)=0,\nonumber\\
    \Rightarrow&2\sum_jp_j\left(\partial_{\Lambda}\text{ln}D+\frac{j}{\xi^2}\partial_{\Lambda}\xi\right)=0,\nonumber\\
    \Rightarrow&\partial_{\Lambda}\text{ln}D=-\frac{\langle j\rangle}{\xi^2}\partial_{\Lambda}\xi,
\end{align}
where $\langle j\rangle$ is defined as the expectation value of the position of the state, namely $\langle j\rangle\equiv\sum_j jp_j$. Similarly, we have $\langle j^2\rangle\equiv \sum_jj^2p_j$ and $\text{Var}(j)\equiv \langle j^2\rangle - \langle j\rangle^2$. Thus, the first term in the Eq.~(\ref{Eq:QFI_pure_state}) is simplified as
\begin{align}
    \langle\partial_{\Lambda}\psi|\partial_{\Lambda}\psi\rangle&=\sum_j p_j \left(\partial_{\Lambda}\text{ln}D+\frac{j}{\xi^2}\partial_{\Lambda}\xi\right)^2+(\partial_{\Lambda}k)^2\sum_j j^2p_j,\nonumber\\
    &=\sum_jp_j\left[(\partial_{\Lambda}\text{ln}D)^2+\frac{2j}{\xi^2}\partial_{\Lambda}\text{ln}D\cdot \partial_{\Lambda}\xi+\left(\frac{j}{\xi^2}\partial_{\Lambda}\xi\right)^2\right]+(\partial_{\Lambda}k)^2\sum_jj^2p_j,\nonumber\\
    &=(\partial_{\Lambda}\text{ln}D)^2+2\partial_{\Lambda}\text{ln}D\cdot \frac{\partial_{\Lambda}\xi}{\xi^2}\langle j\rangle + \left(\frac{\partial_{u}\xi}{\xi^2}\right)^2\langle j^2\rangle + (\partial_{\Lambda}k)^2\langle j^2\rangle,\nonumber\\
    &=\left(\frac{\partial_{\Lambda}\xi}{\xi^2}\right)^2\text{Var}(j)+(\partial_{\Lambda}k)^2\langle j^2\rangle.
\label{Eq:first_term}
\end{align}
Next, we show the algebraic relation between variance $\text{Var}(j)$ and the localization length $\xi$. From the definition of $p_j$, we have $p_j=|D|^2 j^{2p-2}e^{-2j/\xi}$ and redefine the expectation values as
\begin{equation}
    \langle j\rangle=\frac{\sum_j jp_j}{\sum_jp_j}=\frac{\sum_j j^{2p-1}e^{-2j/\xi}}{\sum_jj^{2p-2}e^{-2j/\xi}},\quad \quad \langle j^2\rangle=\frac{\sum_j j^2p_j}{\sum_jp_j}=\frac{\sum_j j^{2p}e^{-2j/\xi}}{\sum_jj^{2p-2}e^{-2j/\xi}}.
\end{equation}
We approach $\Lambda_c$ from the topological phase where a left-localized edge state exists. Therefore, we work in the scaling window $1\ll\xi\ll L$ by fixing $\Lambda_c$ and taking $L\rightarrow \infty$. In this window, the right boundary contributes a negligible $O(e^{-L/\xi})$ weight, and the probability distribution after summation $\sim j^{2p-2}e^{-2j/\xi}$ can be approximated through a continuous integral as
\begin{align}
    &\sum_{j=1}^{\infty}j^{2p-2}e^{-2j/\xi}\approx \int_0^{\infty} j^{2p-2}e^{-2j/\xi}=\frac{\Gamma(2p-1)}{(\frac{2}{\xi})^{2p-1}},\\
    &\sum_{j=1}^{\infty}j^{2p-1}e^{-2j/\xi}\approx \int_0^{\infty} j^{2p-1}e^{-2j/\xi}=\frac{\Gamma(2p)}{(\frac{2}{\xi})^{2p}},\\
    &\sum_{j=1}^{\infty}j^{2p}e^{-2j/\xi}\approx \int_0^{\infty} j^{2p}e^{-2j/\xi}=\frac{\Gamma(2p+1)}{(\frac{2}{\xi})^{2p+1}},
\end{align}
where the gamma function is defined as $\Gamma(z)\equiv\int_0^{\infty}t^{z-1}e^{-t}dt$, with $\Gamma(n)=(n-1)!$. Therefore, the expectation values are simplified by substituting the $\Gamma$ function as
\begin{align}
    &\langle j \rangle\approx \frac{\Gamma(2p)}{\Gamma(2p-1)}\frac{(\frac{2}{\xi})^{2p-1}}{(\frac{2}{\xi})^{2p}}=(2p-1)\frac{\xi}{2},\\
    &\langle j^2\rangle\approx \frac{\Gamma(2p+1)}{\Gamma(2p-1)}\frac{(\frac{2}{\xi})^{2p-1}}{(\frac{2}{\xi})^{2p+1}}=(2p)(2p-1)\left(\frac{\xi}{2}\right)^2,\\
    &\text{Var}(j)= \langle j^2\rangle - \langle j\rangle^2\approx (2p-1)\left(\frac{\xi}{2}\right)^2.
\label{Eq:Variance_dependence}
\end{align}
For $\Lambda\rightarrow \Lambda_c$, the momentum at such limit is $\lim_{\Lambda\rightarrow \Lambda_c}k=k_c$. Hence, the second term in Eq.~(\ref{Eq:QFI_pure_state}) is vanishing at the transition point explicitly
\begin{equation}
    \langle \psi|\partial_{\Lambda}\psi\rangle= \sum_jp_j(\partial_{\Lambda}\text{ln}D+\frac{j}{\xi^2}\partial_{\Lambda}\xi)+i\partial_{\Lambda}k\sum_jjp_j=i\partial_{\Lambda}k\cdot\langle j\rangle=0.
\label{Eq:second_term}
\end{equation}
Substituting Eqs.~(\ref{Eq:first_term}), (\ref{Eq:Variance_dependence}), and (\ref{Eq:second_term}) into Eq.~(\ref{Eq:QFI_pure_state}) yields a compact form of the QFI in terms of the localization length as 
\begin{equation}
    \mathcal{F}_{Q}=4\left(\frac{\partial_{\Lambda}\xi}{\xi^2}\right)^2\text{Var}(j)\sim (2p-1)\left(\frac{\partial_{\Lambda}\xi}{\xi}\right)^2.
\end{equation}
At criticality, the localization length in a finite system would be comparable to the system size, i.e., $\xi\sim L$~\cite{rams2018limits}. Using the algebraic behavior near topological criticality $\xi \sim |\Lambda-\Lambda_c|^{-\frac{1}{p}},\quad \partial_{\Lambda}\xi\sim -\frac{1}{p}|\Lambda-\Lambda_c|^{-\frac{1}{p}-1}=-\frac{1}{p}\xi^{p+1}$, we then recover the QFI scaling
\begin{equation}
    \mathcal{F}_{Q}\sim (2p-1)\left(\frac{-\frac{1}{p}\xi^{p+1}}{\xi}\right)^2\sim \xi^{2p}\sim L^{2p}.
\end{equation}
Note that the multiplicity $p$ of solutions to $P(z\rightarrow z_*,\Lambda)=0$ satisfying $H|\psi_L\rangle=0$ is nothing but the band-touching order of the energy gap in the momentum space. Therefore, near-degenerate $z_j$'s fall altogether on a unit circle when the system is tuned to the critical point $(\Lambda_c,k_c)$, and thus the exponential decay factor $\sim e^{-j/\xi}$ vanishes.

\appsection{\MakeUppercase{Two-dimensional higher-order topological insulators and Chern insulators as probes}}\label{App:2D}
In this section, we generalize the results of the one-dimensional eSSH model discussed in the main text to two-dimensional (2D) systems, further verifying the connection between band touching and probe sensitivity. We first investigate a higher-order topological insulator (HOTI) with long-range hopping~\cite{PhysRevLett.128.127601,PhysRevLett.131.157201}. The Hamiltonian of this model can be written as
\begin{align}
    H_\mathrm{HOTI}(k)
    = &[\lambda_0 + \lambda_1\cos(k_x) + \lambda_2\cos(2k_x)]\Gamma_4 + [\lambda_1\sin(k_x) + \lambda_2\sin(2k_x)]\Gamma_3 \notag \\
    & [\lambda_0 + \lambda_1\cos(k_y) + \lambda_2\cos(2k_y)]\Gamma_2 + [\lambda_1\sin(k_y) + \lambda_2\sin(2k_y)]\Gamma_1,
\label{Eq:HOTI}
\end{align}
where $\lambda_0$ is the intracell hopping strength while $\lambda_{r\neq0}$ denotes the intercell hopping strength between $j$-th unit cell and its $r$-th neighboring unit cells. $\Gamma_1 = -\tau_2\sigma_1, \Gamma_2 = -\tau_2\sigma_2, \Gamma_3 = -\tau_2\sigma_3, \Gamma_4 = \tau_1\sigma_0$ and $\sigma,\tau$ are Pauli
matrices for the degrees of freedom within a unit cell.

Since the model preserves chiral symmetry, its topological properties can be described by the multipole chiral number (MCN), which is a real-space topological invariant~\cite{PhysRevLett.128.127601,PhysRevLett.131.157201}. The MCN can be defined as
\begin{equation}
    N_{xy} = \frac{1}{2\pi i} \mathrm{Tr} \log\!\left( \bar{Q}^{A}_{xy} \bar{Q}^{B\dagger}_{xy} \right) \in \mathbb{Z},
\label{Eq:Nxy}
\end{equation}
where $\bar{Q}^{S}_{xy} = U_{S}^{\dagger} Q^{S}_{xy} U_{S}$, for $S = A,B$, is the sublattice multipole moment operators projected into the spaces $U_{S}$. For the 2D system, the sublattice multipole moment operator $Q_{xy}^{S} = \sum_{\mathbf{R},\, \alpha \in S} \ket{\mathbf{R}, \alpha}\exp\!\left(- i \frac{2\pi xy}{L_x L_y}\right)\bra{\mathbf{R}, \alpha}$. Here, $\mathbf{R}$ denotes the unit cell of the finite system, and we set the system size to satisfy $L_x = L_y = L$. The phase diagram can be obtained from the calculation of the MCN, as shown in Fig.~\ref{fig:Fig.S1}(a).

\begin{figure*}
\includegraphics[scale=0.22]{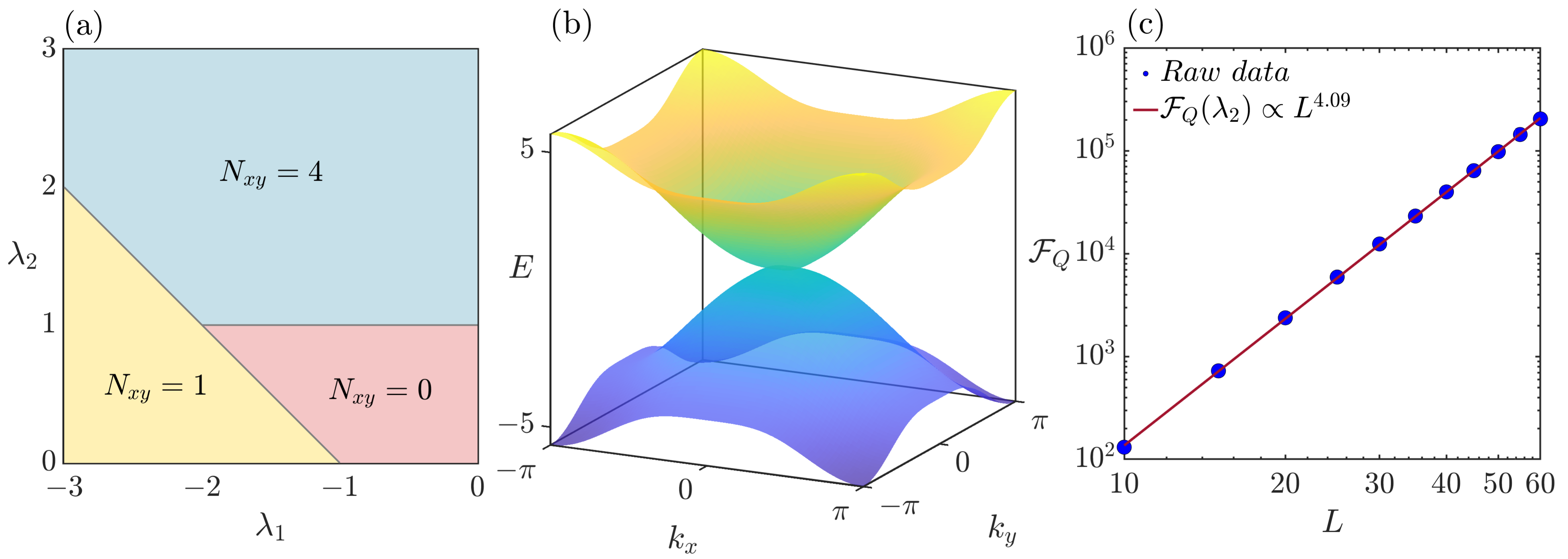}
\caption{
(a) Phase diagram characterized by multipole chiral number $N_{xy}$ of the HOTI model with $\lambda_0=1$. (b) Band structure of the model as a function of momenta $k_x$ and $k_y$ with $\lambda_1=-2$ and $\lambda_2=1$. (c) QFI as a function of system size $L$. The fitting function exhibits a power-law scaling as $\mathcal{F}_{Q}\sim L^{4.09}$, where the blue dots represent numerical simulation results and the red line represents the polynomial fitting function.}
\label{fig:Fig.S1}
\end{figure*}

A nonzero $N_{xy}$ indicates that the system is in a higher-order topological phase. Due to the symmetry $C_4$ preserved by the model, the total number of higher-order topological states equals four times $N_{xy}$. The boundaries between different phases correspond to topological phase transitions and band touching. The results in the main text indicate that long-range hopping can give rise to high-order band-touching, thereby enhancing the scaling behavior of the QFI. In this HOTI model, we introduce the next-nearest-neighbor hopping $\lambda_2$, which gives rise to a second-order band-touching. The energies of $H_\mathrm{HOTI}(k)$ can be represented as
\begin{equation}
    E_\mathrm{HOTI}(k) =\pm\sqrt{d_1^2(k_y)+d_2^2(k_y)+d_3^2(k_x)+d_4^2(k_x)},
\label{Eq:E_HOTI}
\end{equation}
with 
\begin{align}
    d_1(k_y) &= \lambda_1 \sin k_y + \lambda_2 \sin(2k_y),\\
    d_2(k_y) &= \lambda_0 + \lambda_1 \cos k_y + \lambda_2 \cos(2k_y),\\
    d_3(k_x) &= \lambda_1 \sin k_x + \lambda_2 \sin(2k_x),\\
    d_4(k_x) &= \lambda_0 + \lambda_1 \cos k_x + \lambda_2 \cos(2k_x).
\label{Eq:d_HOTI}
\end{align}

Considering $k_x = k_y = 0$ (i.e., high-symmetry point) in the momentum space, the condition for the second-order band touching can be obtained from Eqs.~(\ref{Eq:E_HOTI})$\--$(\ref{Eq:d_HOTI}) as $\lambda_1 = -2\lambda_2$ and $\lambda_2 = \lambda_0$. We choose the parameters $(\lambda_0, \lambda_1,\lambda_2) = (1, -2, 1)$, which correspond to the multicritical point of the three phase regions shown in Fig.~\ref{fig:Fig.S1}(a). The corresponding band structure is displayed in Fig.~\ref{fig:Fig.S1}(b), where the second-order band touching is confirmed. To verify our scaling relation between band touching and QFI in the HOTI model, we numerically investigate the scaling behavior of QFI. In the calculation of the QFI, we employ the higher-order topological state, i.e., the zero-energy state, with the critical parameter given by $\lambda_{2}=1$. As can be seen from Fig.~\ref{fig:Fig.S1}(c), the QFI satisfies $\mathcal{F}_{Q}\sim L^{2p}$ with $p=2$ for the second-order band touching, which agrees well with our scaling analysis in the main text.

We further investigate the relationship between band touching and quantum metrology in a 2D Chern insulator (CI). We consider a two-band CI~\cite{Bernevig+2013,RevModPhys.95.011002}, with the Hamiltonian given by
\begin{equation}
    H_\mathrm{CI}(k) =d_x(k)\sigma_1+d_y(k)\sigma_2+d_z(k)\sigma_3,
\label{Eq:CI}
\end{equation}
with 
\begin{align}
    &d_x(k)+id_y(k) = [g(k)]^2,\\
    &g(k) = (e^{-ik_x}-1)+i(e^{-ik_y}-1),\\
    &d_z(k) = m_0 + \lambda_0[\cos(k_x)+\cos(k_y)],
\label{Eq:dxdydz}
\end{align}
where $m_0$ and $\lambda_0$ are mass parameters that control the band touching. $d_x(k)$ and $d_y(k)$ contain long-range hoppings. $\sigma_{1,2,3}$ are Pauli
matrices for the degrees of freedom within a unit cell. The topological properties of CI model can be described by the Chern number~\cite{Bernevig+2013,RevModPhys.95.011002}. The phase diagram is shown in Fig.~\ref{fig:Fig.S2}(a). A nonzero Chern number $C$ indicates that the system is in a topological phase. $C=2$ and $C=-2$ correspond to edge states with equal numbers but opposite propagation directions.

\begin{figure*}[b]
\includegraphics[scale=0.22]{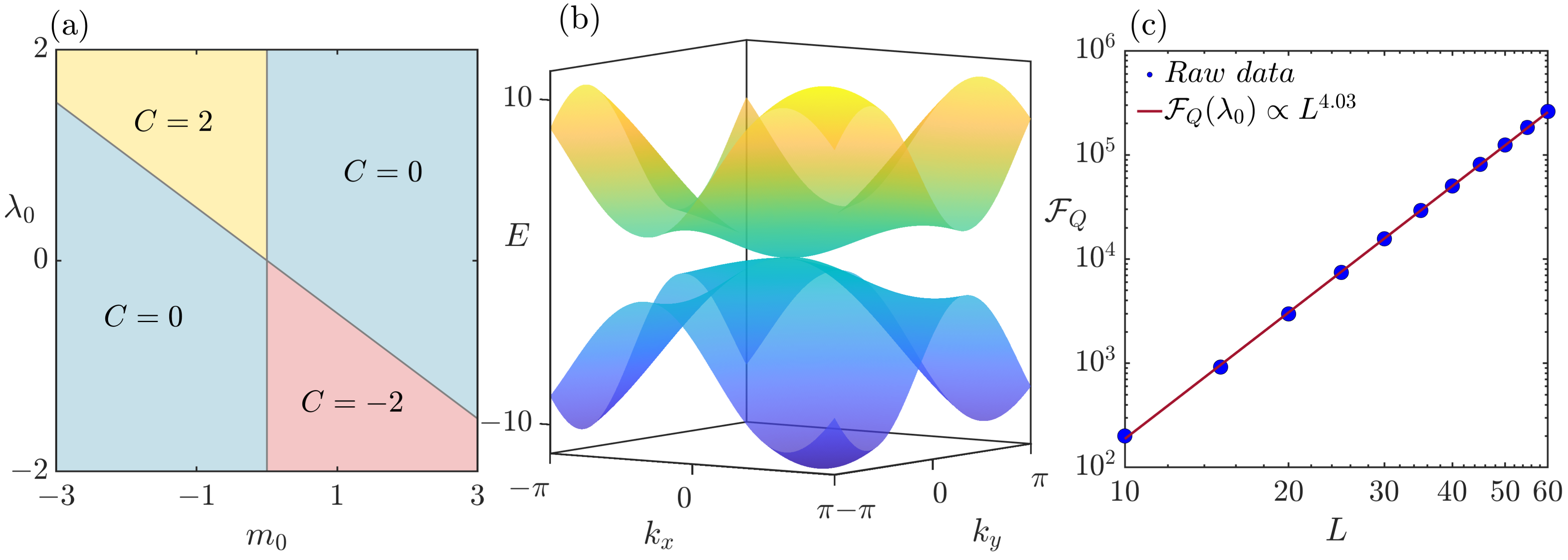}
\caption{
(a) Phase diagram characterized by Chern number $C$ of the CI model. (b) Band structure of the model as a function of momenta $k_x$ and $k_y$ with $m_0=1$ and $\lambda_0=-0.5$. (c) QFI as a function of system size $L$. The fitting function exhibits a power-law scaling as $\mathcal{F}_{Q}\sim L^{4.03}$, where the blue dots represent numerical simulation results and the red line represents the polynomial fitting function.}
\label{fig:Fig.S2}
\end{figure*}

Similar to the HOTI model, the energies of $H_\mathrm{CI}(k)$ can be represented as
\begin{equation}
    E_\mathrm{CI}(k) =\pm\sqrt{d_x^2(k)+d_y^2(k)+d_z^2(k)}.
\label{Eq:E_CI}
\end{equation}
The conditions for band touching are $m_0=-2\lambda_0$ and $m_0=0$, which correspond to the points $(k_x,k_y) = (0,0)$ and $(k_x,k_y) = (-\pi/2,\pi/2)$ in momentum space, respectively. These conditions also define the phase boundaries between different regions in the phase diagram shown in Fig.~\ref{fig:Fig.S2}(a). Unlike the HOTI model, all the band touchings at the phase boundary $m_0=-2\lambda_0$ are of second order. For $m_0=1$ and $\lambda_0=-0.5$, a representative band touching is shown in Fig.~\ref{fig:Fig.S2}(b). However, at the point $(k_x,k_y) = (-\pi/2,\pi/2)$ corresponding to $m_0=0$, the band touching is a mixture of linear and quadratic types, and thus not a purely second-order band touching.

To verify our scaling relation between band touching and QFI in the CI model, we numerically investigate the scaling behavior of QFI based on the band touching shown in Fig.~\ref{fig:Fig.S2}(a). In the QFI calculation, we set the system size to $L_x = L_y = L$. Moreover, we employ the edge states near zero energy, with the critical parameter given by $\lambda_0 =-0.5$. As can be seen from Fig.~\ref{fig:Fig.S2}(c), the QFI satisfies $\mathcal{F}_{Q}\sim L^{2p}$ with $p=2$ for the second-order band touching, which agrees well with our scaling analysis in the main text. In summary, the numerical results of both the HOTI and CI models verify the universality of the connection between band touching and quantum metrology discussed in the main text.

\appsection{\MakeUppercase{Position Basis as Optimal Measurement}}\label{App:Opt_Measure}
In this section, we provide a clean proof of saturating the QFI of the critical edge state using a position measurement. Without the loss of generality, let the left edge state be written in the single-particle position basis as
\begin{equation}
    |\psi_L\rangle=\sum_{j=1}^L \phi_j(\lambda)|j\rangle, \qquad \phi_j(\lambda)=\sqrt{p_j(\lambda)}e^{i\varphi_j(\lambda)},
\end{equation}
where $|j\rangle=a_j^{\dagger}|0\rangle$ is the sublattice-$A$ edge mode. From the pure-state QFI formula, we first expand the state derivative as
\begin{equation}
    |\partial_\lambda\psi\rangle=\sum_j e^{i\varphi_j}\Big(\frac{\partial_\lambda p_j}{2\sqrt{p_j}}+i\sqrt{p_j}\,\partial_\lambda\varphi_j\Big)|j\rangle.
\end{equation}
Therefore,
\begin{align*}
    \langle\partial_\lambda \psi|\partial_\lambda\psi\rangle&=\frac{1}{4}\sum_j\frac{(\partial_\lambda p_j)^2}{p_j}+\sum_jp_j(\partial_\lambda\varphi_j)^2,\\
    \langle\psi|\partial_\lambda\psi\rangle&=\frac{1}{2}\sum_j\partial_\lambda p_j+i\sum_jp_j\partial_\lambda\varphi_j=i\sum_jp_j\partial_\lambda\varphi_j.
\end{align*}
A position measurement uses the POVM $\{\Pi_j=|j\rangle\langle j|\}$ so the probabilities are $p_j(\lambda)=|\phi_j(\lambda)|^2$. Substituting into the pure-state QFI formula gives
\begin{equation}
    \mathcal{F}_Q=\mathcal{F}_C+4\Big[\sum_jp_j(\partial_\lambda\varphi_j)^2-\Big(\sum_jp_j\partial_\lambda\varphi_j\Big)^2\Big]=\mathcal{F}_C+4\text{Var}(\partial_\lambda\varphi_j).
\end{equation}
So the position measurement saturates the QFI iff $\partial_{\lambda}\varphi_j$ is independent of $j$. This is the general saturation criterion in the position basis.

Now we apply this language to the eSSH edge states near the criticality. As discussed in the first section, the edge-state asymptotic profile near the criticality is $\phi_j(\lambda)=D(\lambda)j^{p-1}e^{ik(\lambda)j}e^{-j/\xi(\lambda)}$. Hence, the phase and its derivative are 
\begin{align*}
    \varphi_j&=\text{arg}\,D(\lambda)+k(\lambda)j,\\
    \partial_\lambda\varphi_j&=\partial_\lambda\text{arg}\,D(\lambda)+j\partial_\lambda k(\lambda).
\end{align*}
Note that the term $\partial_\lambda\text{arg}\,D(\lambda)$ is zero in the variance, hence the above identity becomes
\begin{equation}
    \mathcal{F}_Q=\mathcal{F}_C+4(\partial_\lambda k
    )^2\Big[\sum_jp_jj^2-\Big(\sum_jp_jj\Big)^2\Big]=\mathcal{F}_C+4(\partial_\lambda k)^2\text{Var}(j).
\end{equation}
Again, the $\partial_\lambda k$ is vanishing near the critical point $(\lambda_c,k_c)$, i.e., there is no nontrivial parameter-dependent phase $k$ in the edge state amplitudes when measuring the position basis. Therefore, a position measurement saturates the QFI as $\mathcal{F}_Q=\mathcal{F}_C$.

\appsection{\MakeUppercase{QFI scaling through adiabatic process of many-body non-interacting probes}}\label{App:Multi-particle}

In this section, we show that by adiabatically driving a GHZ state, one can actually achieve the hyperscaling of QFI with both system size and particle excitations as $\mathcal{F}_Q\sim N^2L^{2p}$. Let us first consider a single-particle Hamiltonian in Eq.~(\ref{Eq:SM_Hamiltonian_real_space}) as $H=\sum_{r=1}^R\lambda_rH_r$, in which the single parameter $\lambda_r$ governing the hopping strength is of our interest. To introduce the particle dependence in QFI, we prepare our initial state, away from criticality, as a GHZ state~\cite{gio2004}, which is a well-known maximally entangled state as
\begin{equation}
    |\Phi_0\rangle=\frac{|L_1(\lambda(0))\rangle^{\otimes N}+|L_2(\lambda(0))\rangle^{\otimes N}}{\sqrt{2}},
\end{equation}
where $|L_1(\lambda(0))\rangle,|L_2(\lambda(0))\rangle$ are the two left-localized zero-energy edge modes with $\lambda(0)$ away from the critical point at $t=0$, $N$ is the number of particle excitations. We adopt the formula for pure states~\cite{Matteo} $\mathcal{F}_Q=4(\langle\partial_{\lambda}\Phi_{\lambda}|\partial_{\lambda}\Phi_{\lambda}\rangle-|\langle\Phi_{\lambda}|\partial_{\lambda}\Phi_{\lambda}\rangle|^2)$ to evaluate the QFI. After adiabatically driving the two components of the initial GHZ state to the criticality, we yield
\begin{equation}
    |\Phi_{\lambda}\rangle=\frac{|\psi_1(\lambda)\rangle^{\otimes N}+|\psi_2(\lambda)\rangle^{\otimes N}}{\sqrt{2}}=\frac{(U_{\lambda}(T,0)|L_1(\lambda(0))\rangle)^{\otimes N}+(U_{\lambda}(T,0)|L_2(\lambda(0))\rangle)^{\otimes N}}{\sqrt{2}},
\end{equation}
where $U_\lambda(T,0)=\mathcal{T}\text{e}^{-i\int_0^{T}H(t')dt'}$ is the adiabatic time evolution unitary operator with $\mathcal{T}$ the time-ordering operator, $T$ is the adiabatic evolution time which scales as $T\sim L^z$ in general with $z$ the dynamical critical exponent. In the following, we consider two different cases with and without the finite-size effect.
As mentioned in the main text, in the thermodynamic limit, the edge-state subspace is exactly degenerate, and the zero modes localized at sublattice $A$ never overlap with those localized at sublattice $B$ before reaching the transition point $\lambda_c$. Therefore, no non-adiabatic mixing is allowed in this limit, and the parameterized GHZ is written as
\begin{equation}
    |\Phi_{\lambda}\rangle=\frac{|L_1(\lambda)\rangle^{\otimes N}+e^{iN\phi(\lambda)}|L_2(\lambda)\rangle^{\otimes N}}{\sqrt{2}}=\frac{|A(\lambda)\rangle+e^{iN\phi(\lambda)}|B(\lambda)\rangle}{\sqrt{2}},
\end{equation}
where $\phi$ is the relative phase of the single-particle eigenstate accumulated during the adiabatic process, composed of dynamical phase $\theta$ and geometric phase $\gamma$. The geometric phase $\gamma$ can be chosen as zero because the geometric phase $\gamma$ itself does not accumulate over the adiabatic evolution time and hence does not contribute to any additional scaling to QFI. For simplicity, we do not bother writing the $\lambda$-dependence explicitly, and choose the parallel-transport gauge during the adiabatic process so that $\langle L_1|\partial_{\lambda}L_1\rangle=\langle L_2|\partial_{\lambda}L_2\rangle=0$. Differentiating the two GHZ components $|A\rangle,|B\rangle$ as well as the final state $|\Phi_{\lambda}\rangle$ gives
\begin{align}
    |\partial_{\lambda}A\rangle&=\sum_{j=1}^N|L_1\rangle^{\otimes j-1}\otimes |\partial_{\lambda}L_1\rangle\otimes |L_1\rangle^{\otimes N-j},\\
    |\partial_{\lambda}B\rangle&=\sum_{j=1}^N|L_2\rangle^{\otimes j-1}\otimes |\partial_{\lambda}L_2\rangle\otimes |L_2\rangle^{\otimes N-j},\\
    |\partial_{\lambda}\Phi\rangle&=\frac{1}{\sqrt{2}}[|\partial_{\lambda}A\rangle+iN(\partial_{\lambda}\phi)e^{iN\phi}|B\rangle+e^{iN\phi}|\partial_{\lambda}B\rangle].
\end{align}
Note that due to our choice of gauge and orthogonality, the cross terms in the inner product $\langle\partial_{\lambda}\Phi|\partial_{\lambda}\Phi\rangle$ vanish. Let us consider the first-order cross terms
\begin{align}
    \langle A|\partial_{\lambda}A\rangle&=\sum_{j=1}^N \langle L_1|L_1\rangle^{j-1}\langle L_1|\partial_{\lambda}L_1\rangle\langle L_1|L_1\rangle^{N-j}=0,\\
    \langle B|\partial_{\lambda}B\rangle&=\sum_{j=1}^N \langle L_2|L_1\rangle^{j-1}\langle L_2|\partial_{\lambda}L_2\rangle\langle L_2|L_2\rangle^{N-j}=0,\\
    \langle A|\partial_{\lambda}B\rangle&=\sum_{j=1}^N \langle L_1|L_2\rangle^{j-1}\langle L_1|\partial_{\lambda}L_2\rangle\langle L_1|L_2\rangle^{N-j}=0,\\
    \langle B|\partial_{\lambda}A\rangle&=\sum_{j=1}^N \langle L_2|L_1\rangle^{j-1}\langle L_2|\partial_{\lambda}L_1\rangle\langle L_2|L_1\rangle^{N-j}=0.
\end{align}
Particularly, the second-order cross term also vanishes as:\\ for $j_1>j_2$:
\begin{equation}
    \langle \partial_{\lambda}A|\partial_{\lambda}B\rangle=\frac{1}{2}\sum_{j_1,j_2=1}^N \langle L_1|L_2\rangle^{j_2-1}(\langle L_1|^{\otimes j_1-j_2}\otimes \langle \partial_{\lambda}L_1|\otimes \langle L_1|^{\otimes N-j_1})(|\partial_{\lambda}L_2\rangle\otimes |L_2\rangle^{\otimes N-j_2})=0,
\end{equation}
for $j_1=j_2$:
\begin{equation}
    \langle \partial_{\lambda}A|\partial_{\lambda}B\rangle=\sum_{j_1=1}^N \langle L_1|L_2\rangle^{j_1-1}\langle \partial_{\lambda}L_1|\partial_{\lambda}L_2\rangle\langle L_1|L_2\rangle^{N-j_1}=0,
\end{equation}
for $j_1<j_2$:
\begin{equation}
    \langle \partial_{\lambda}A|\partial_{\lambda}B\rangle=\frac{1}{2}\sum_{j_1,j_2=1}^N \langle L_1|L_2\rangle^{j_1-1}(\langle\partial_{\lambda}L_1|\otimes \langle L_1|^{\otimes N-j_1})(|L_1\rangle^{\otimes j_2-j_1}\otimes |\partial_{\lambda}L_2\rangle\otimes |L_2\rangle^{\otimes N-j_2})=0.
\end{equation}
Therefore, the second-order cross terms vanish as $\langle\partial_{\lambda}A|\partial_{\lambda}B\rangle=\langle\partial_{\lambda}B|\partial_{\lambda}A\rangle=0$. Diagonal terms contribute the $\sim N$ dependence as 
\begin{align}
    \langle \partial_{\lambda}A|\partial_{\lambda}A\rangle&=\sum_{j_1,j_2=1}^N (\langle L_1|^{\otimes j_1-1}\otimes \langle \partial_{\lambda}L_1|\otimes \langle L_1|^{\otimes N-j_1})(|L_2\rangle^{\otimes j_2-1}\otimes |\partial_{\lambda}L_2\rangle \otimes |L_2\rangle^{\otimes N-j_2})\nonumber\\
    &=\sum_{j=1}^N \langle \partial_{\lambda}L_1|\partial_{\lambda}L_1\rangle+\sum_{j_1\neq j_2,j_1,j_2=1}^N \langle \partial_{\lambda}L_1|L_1\rangle \prod_{j_3\neq j_1,j_2}\langle L_1|L_1\rangle \langle \partial_{\lambda} L_1|L_1\rangle\nonumber\\
    &=N\langle\partial_{\lambda}L_1|\partial_{\lambda}L_1\rangle.
\end{align}
Similarly, $\langle \partial_{\lambda}B|\partial_{\lambda}B\rangle=N\langle\partial_{\lambda}L_2|\partial_{\lambda}L_2\rangle$. Hence, the inner products become
\begin{align}
    \langle\partial_{\lambda}\Phi|\partial_{\lambda}\Phi\rangle&=\frac{1}{2}[\langle\partial_{\lambda}A|\partial_{\lambda}A\rangle+N^2(\partial_{\lambda}\phi)^2+\langle\partial_{\lambda}B|\partial_{\lambda}B\rangle],\\
    \langle\Phi|\partial_{\lambda}\Phi\rangle&=\frac{1}{2}[\langle A|\partial_{\lambda}A\rangle+iN(\partial_{\lambda}\phi)e^{iN\phi}\langle A|B\rangle+e^{iN\phi}\langle A|\partial_{\lambda}B\rangle+e^{-iN\phi}\langle B|\partial_{\lambda}A\rangle+iN(\partial_{\lambda}\phi)+\langle B|\partial_{\lambda}{B\rangle}]=\frac{i}{2}N(\partial_{\lambda}\phi),\\
    |\langle\Phi|\partial_{\lambda}\Phi\rangle|^2&=\frac{1}{4}N^2(\partial_{\lambda}\phi)^2.
\end{align}
The QFI for the final state $|\Phi_{\lambda}\rangle$ with respect to the parameter $\lambda$ is
\begin{equation}
    \mathcal{F}_Q(\lambda)=4[\langle\partial_{\lambda}\Phi_{\lambda}|\partial_{\lambda}\Phi_{\lambda}\rangle-|\langle\Phi_{\lambda}|\partial_{\lambda}\Phi_{\lambda}\rangle|^2]=\mathcal{F}_Q^{\text{int}}(\lambda)+\mathcal{F}_Q^{\text{eig}}(\lambda)=N^2(\partial_{\lambda}\phi)^2+2N(\langle\partial_{\lambda} L_1|\partial_{\lambda}L_1\rangle+\langle\partial_{\lambda} L_2|\partial_{\lambda}L_2\rangle),
\end{equation}
which is regarded as the addition of contributions from both the relative phase (interference term) $\mathcal{F}_Q^{\text{int}}$ and from the eigenstate gap closing $\mathcal{F}_Q^{\text{eig}}$. Let us first simplify $\mathcal{F}_Q^{\text{eig}}$ and check how it scales with system size $L$ and number of excitations $N$. Straightforwardly, the inner products of the partial derivative of edge modes can be simplified through first-order perturbation theory
\begin{align}
    \langle\partial_{\lambda}L_1|\partial_{\lambda}L_1\rangle=\sum_{n\neq 1}\frac{|\langle L_n|H_r|L_1\rangle|^2}{(E_n-E_1)^2},\quad \langle\partial_{\lambda}L_2|\partial_{\lambda}L_2\rangle=\sum_{n\neq 2}\frac{|\langle L_n|H_r|L_2\rangle|^2}{(E_n-E_2)^2},
\end{align}
where $E_1$ and $E_2$ represent the eigenvalues with respect to $|L_1\rangle$ and $|L_2\rangle$. Recalling the main argument of this letter $\Delta E\sim L^{-p}$ with $p$ being the order of band touching, we have
\begin{equation}
    \mathcal{F}_Q^{\text{eig}}(\lambda)=2N\left[\sum_{n\neq 1}\frac{|\langle L_n|H_r|L_1\rangle|^2}{(E_n-E_1)^2}+\sum_{n\neq 2}\frac{|\langle L_n|H_r|L_2\rangle|^2}{(E_n-E_2)^2}\right]\sim NL^{2p}.
\end{equation}

\begin{figure*}
  \includegraphics[scale=0.25]{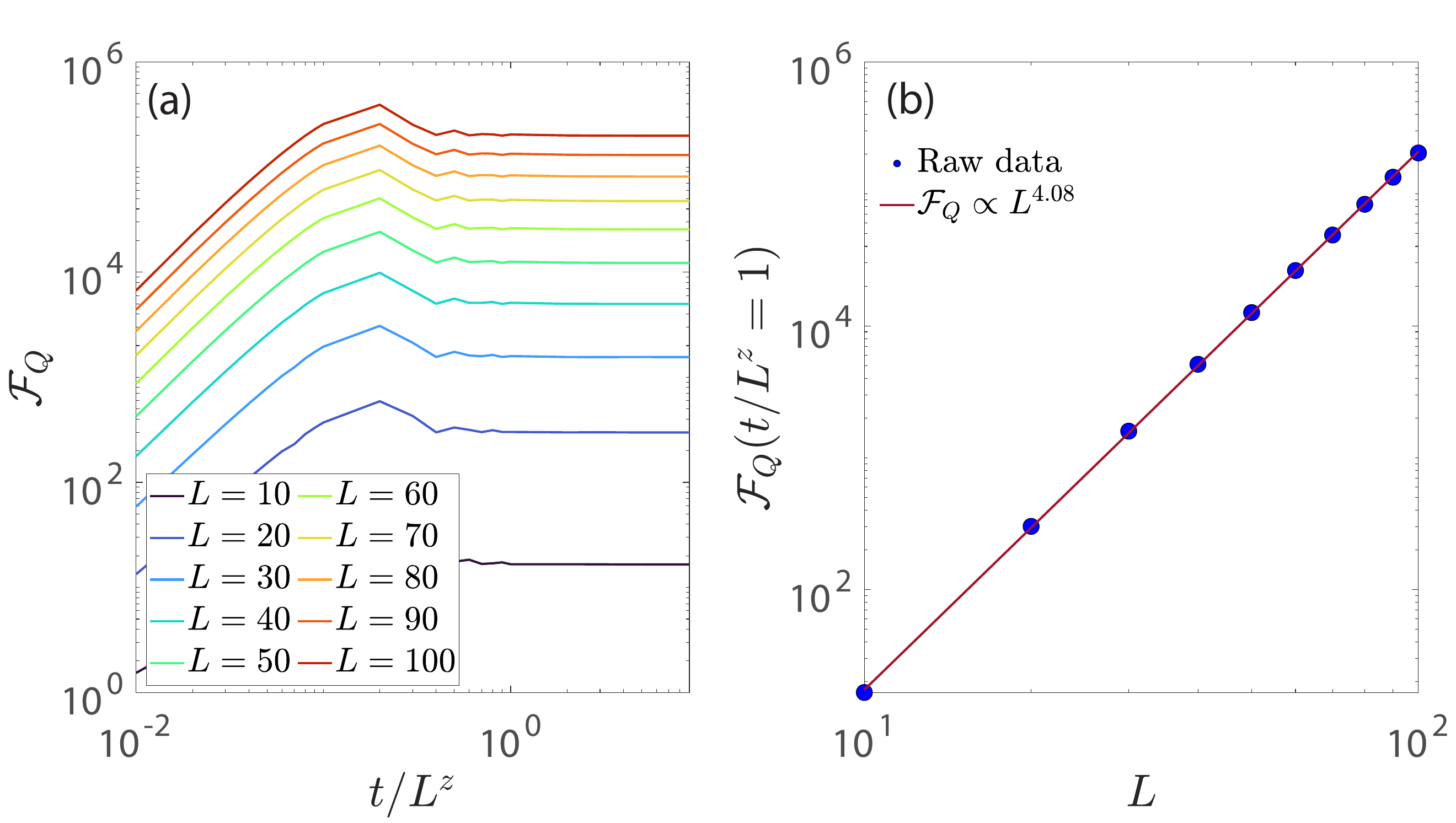}
  \caption{The QFI as a function of rescaled total evolution time $T/L^z$ for the eSSH model at the phase transition point $(\lambda_0,\lambda_1,\lambda_2)=(1,2,1)$. The deviation in the parameter away from the $\lambda_c$ is set as $\delta\lambda=10^{-10}$.}
  \label{fig:comp}
\end{figure*}

Secondly, we show that the interference term actually contributes to the $\sim N^2L^{2p}$ to the QFI. Since in our extended SSH model the time-reversal symmetry is conserved, and thus the eigenstates are all real functions, the geometric phase is always vanishing. The only contribution to the phase difference is from the relative dynamical phase $\theta(\lambda)=-\int_0^T[E_2(t')-E_1(t')]dt'$. In addition, recall the standard finite-size arguments~\cite{rams2018limits} of localization length near the phase transition points as $\xi\sim L\sim|\lambda-\lambda_c|^{-\nu},\Delta E\sim L^{-z}\sim |\lambda-\lambda_c|^{\nu z}$, where $\nu$ is the critical exponent controlling the localization length divergence. The interference term becomes $\mathcal{F}_Q^{\text{int}}=N^2(\partial_{\lambda}\theta)^2=N^2(\int_0^Tdt'\partial_{\lambda}\Delta E)^2\sim N^2(\int_0^Tdt'\partial_{\lambda}|\lambda-\lambda_c|^{\nu z})^2=N^2T^2=N^2L^{2p}$, where we used $\nu=1/p=1/z$ proven in Eq.~(\ref{Eq:localization_length_dlambda}). Notably, the reason why we replace ``$T\gg\Delta E^{-1}$'' by ``$T\sim\Delta E^{-1}$'' is that from both the results from Ref.~\cite{rams2018limits} and numerical simulations of our model in Fig.~\ref{fig:comp}, the total evolution time $T\sim 1/\Delta$ is enough for us to see the asymptotic scaling of the QFI. A longer timescale would not provide any new extractable information from the dynamics. Again, by the same token of taking the derivative of the time-evolved state, it is apparent that the geometric phase is independent of adiabatic evolution time and hence contributes no additional scaling to QFI. Therefore, the total QFI accumulated in the noninteracting $N$-particle adiabatic time evolution is
\begin{equation}
    \mathcal{F}_Q(\lambda)=N^2(\partial_{\lambda}\theta)^2+2N\left[\sum_{n\neq 1}\frac{|\langle L_n|H_r|L_1\rangle|^2}{(E_n-E_1)^2}+\sum_{n\neq 2}\frac{|\langle L_n|H_r|L_2\rangle|^2}{(E_n-E_2)^2}\right].
\end{equation}

\appsection{\MakeUppercase{Variation of Critical Exponents near the Multi-criticality}}\label{App:Multicriticality}
Here, we derive the various values of critical exponents $\nu$ near the quartic touching point $(\lambda_0,\lambda_1,\lambda_2,\lambda_3,\lambda_4)=(1,-4,6,-4,1)$ of the eSSH model with $R=4$, through the renormalization group (RG) flow analysis. Let us write the eSSH Bloch Hamiltonian as
\begin{equation}
    H(k)=\begin{pmatrix}
        0&h(k)\\h^*(k)&0
    \end{pmatrix},\qquad h(k)=\sum_r\lambda_re^{ikr}.
\end{equation}
At the quartic touching point, $h(k)=e^{4ik}-4e^{3ik}+6e^{2ik}-4e^{ik}+1$, so the gap closes at $k_c=0$ and $E(k)\sim |k-k_c|^4$. Without the loss of generality, set $\lambda_0=1$ as a unit of the other parameters and perturb around this point by rewriting them
\begin{equation}
    \lambda_1=-4+\delta_1,\qquad \lambda_2=6+\delta_2,\qquad \lambda_3=-4+\delta_3,\qquad \lambda_4=1+\delta_4,
\end{equation}
and expanding the energy near the critical momentum $|k-k_c|=q\ll 1$ gives
\begin{equation}
    h(q)=u_0+iu_1q+u_2q^2+iu_3q^3+C_4q^4+O(q^5),
    \label{Eq:low-energy}
\end{equation}
with
\begin{align*}
    u_0&=\delta_1+\delta_2+\delta_3+\delta_4,\\
    u_1&=\delta_1+2\delta_2+3\delta_3+4\delta_4,\\
    u_2&=-\frac{\delta_1}{2}-2\delta_2-\frac{9}{2}\delta_3-8\delta_4,\\
    u_3&=-\frac{\delta_1}{6}-\frac{4}{3}\delta_2-\frac{9}{2}\delta_3-\frac{32}{3}\delta_4,
\end{align*}
and
\begin{equation}
    C_4=1+O(\delta).
\end{equation}
Since $C_4\rightarrow1\neq 0$ at the fixed point, we can renormalize it to 1 and keep
\begin{equation}
    h(q)\simeq u_0+iu_1q+u_2q^2+iu_3q^3+q^4,
\end{equation}
Under Wilson rescaling in the real configuration space $x\rightarrow x/b$ with $b\ge 1$, the equivalent rescaling of the momentum $q\rightarrow bq$
already keeps the low-energy effective theory invariant. Therefore, the perturbations have to scale by a simple power-law prefactor
\begin{equation}
    u_0\rightarrow b^4u_0,\qquad u_1\rightarrow b^3u_1,\qquad u_2\rightarrow b^2u_2,\qquad u_3\rightarrow bu_3.
\end{equation}
If we define the RG rescaling coordinate as $l\equiv\text{ln}\,b$, the RG equations are
\begin{equation}
    \frac{du_0}{dl}=4u_0,\qquad \frac{du_1}{dl}=3u_1,\qquad \frac{du_2}{dl}=2u_2,\qquad \frac{du_3}{dl}=u_3.
\end{equation}
So the quartic fixed point has four unstable directions with RG eigenvalues
\begin{equation}
    y_0=4,\qquad y_1=3,\qquad y_2=2,\qquad y_3=1.
\end{equation}
The solutions to the RG equations are
\begin{equation}
    u_m(l)=u_m(0)e^{y_ml}=u_m(0)e^{(4-m)l},\qquad m=0,1,2,3.
\end{equation}
Now we extract the critical exponent. Suppose we approach the quartic touching point along a one-parameter path by varying $\lambda$ so that the perturbations are linear in $\lambda$; then $u_m=\alpha_m\lambda$.
If $u_m$ is the leading term of the nonvanishing perturbation, the RG flow approaches the fixed point when
\begin{equation}
    u_m(l_*)\sim 1\quad \Rightarrow\quad e^{(4-m)l_*}|\lambda|\sim 1,
\end{equation}
where $l_*$ is the crossover RG rescaling factor beyond which the system no longer looks critical. Under RG the physical length shrinks as $\xi(l)\sim \xi e^{-l}$, thus at the exact fixed point the rescaled correlation length is just order one $\xi(l_*)\sim O(1)$ and hence
\begin{equation}
   \xi\sim e^{l_*}\sim |\lambda|^{-\frac{1}{4-m}}.
\end{equation}
So the critical exponent controlling the correlation length divergence is
\begin{equation}
    \nu=\frac{1}{4-m}=\frac{1}{y_m}.
\end{equation}
From Eq.~(\ref{Eq:low-energy}), along the path approaching the quartic touching point, where the lowest order of the nonvanishing field is $M$, controls the actual critical exponent $\nu=\frac{1}{y_M}$. For instance, the path with $u_0=0,\ u_1\neq0$ gives the $\nu=\frac{1}{3}$ and $u_0=u_1=0,\ u_2\neq 0$ gives $\nu=\frac{1}{2}$. A general approaching path with $u_0\neq 0$ gives $\nu=\frac{1}{4}$. Therefore, the critical exponent $\nu$ changes only when the path is special so that the relevant scaling fields vanish.

\makeatletter
\setbox\widetext@bot=\hbox{}
\makeatother
\end{widetext}

\bibliography{Refs}

\end{document}